\documentclass[twocolumn,showpacs,aps,prd,superscriptaddress]{revtex4}
\usepackage{graphicx}
\usepackage{dcolumn}
\usepackage{amsmath}
\usepackage{epsfig}
\usepackage{colordvi}
\usepackage{color}
\usepackage{hhline}

\begin{document}

\title{\bf\boldmath 
Evidence for the decay $\psi(3770)\to K^+K^-$}
\author{V.P.~Druzhinin}
\affiliation{Budker Institute of Nuclear Physics, SB RAS, Novosibirsk, 630090, Russia}
\affiliation{Novosibirsk State University, Novosibirsk 630090, Russia}

\begin{abstract}
Existing data on the $e^+e^-\to K^+K^-$ cross section
at the center-of-mass energy above 2.6 GeV
are fitted with a sum of $\psi(3770)$ resonant and 
continuum contributions. Two solutions for the resonance
production cross section are found with a significance 
of 3.2$\sigma$.
Data on the $e^+e^-\to K_SK_L$ cross section are used to resolve 
the ambiguity and
for further constraining the values of the 
$e^+e^-\to \psi(3770) \to K^+K^-$ cross section 
and the interference phase. They are found to be
$\sigma_{\psi(3770)}=0.073^{+0.061}_{-0.044}$ pb 
and $\phi=(309^{+17}_{-35})^\circ$, respectively. 
The same fitting procedure for the $\psi(4160)$ resonance
leads to the upper limit on the $e^+e^-\to \psi(4160) \to K^+K^-$
cross section $\sigma_{\psi(4160)}<0.062$ pb at 90\% confidence
level.
\end{abstract}
\pacs{13.25.Gv,13.66.Bc,14.40.Pq}

\maketitle

\section{Introduction}
The $\psi(3770)$ meson is the lowest-mass $c\bar{c}$ state 
laying above the open-charm threshold and therefore is expected to decay
predominantly into $D\bar{D}$ pairs. A simple estimation based on the
assumption that the decay probability for $c\bar{c}$-meson
into light hadrons is proportional to $|\Psi(0)|^2$, 
where $\Psi(0)$ is the meson wave function at the origin, leads 
to the relation between $\psi(3770)$ and $\psi(2S)$ branching fractions:
\begin{eqnarray}
B(\psi(3770)\to f)&\approx& B(\psi(2S)\to f)
\frac{B(\psi(3770)\to e^+e^-)}{B(\psi(2S)\to e^+e^-)}\nonumber\\
&\approx&10^{-3}B(\psi(2S)\to f)
\label{eq0}
\end{eqnarray}
The total branching fraction of the $\psi(3770)$ into light hadrons is expected
to be about $10^{-3}$. This prediction strongly contradicts 
the BES Collaboration observation that the total branching fraction
for non-$D\bar{D}$ decays is $(14.5\pm1.7\pm5.8)\%$~\cite{nonDD}.
The CLEO measurement of the same value is 
$B(\psi(3770)\to\mbox{non-}D\bar{D}) < 9\%$ at 90\% confidence 
level (CL)~\cite{nonDD1}. The BES measurement has triggered 
an intensive search for $\psi(3770)$ decays into light hadrons.
About 90 final states were studied~\cite{pdg}, but
only two decays, to $\phi\eta$~\cite{phieta} and $p\bar{p}$~\cite{ppbar},
were observed.
The measured branching fractions significantly exceed above prediction,
by more than an order of magnitude for $\psi(3770)\to p\bar{p}$ and by four 
orders for $\psi(3770)\to\phi\eta$ decay.
The mechanism explaining relatively large values of branching fractions
is production of light hadrons via intermediate $D\bar{D}$ loops 
(see Refs.~\cite{ddloop,th2} and references therein). 
The predicted in Ref.~\cite{ddloop} branching fraction for the decay 
$\psi(3770)\to K^+K^-$, studied in this work, is  $9\times10^{-5}$; it is larger
than the prediction of Eq.~(\ref{eq0}) by about three orders of magnitude.

For $\psi(3770)$ branching fractions of $10^{-4}$, the cross section
for the resonant process $e^+e^-\to \psi(3770) \to f$ is usually less than the 
nonresonant $e^+e^- \to f$ cross section. Therefore, the $\psi(3770)$ decay 
will reveal itself as an interference pattern in the energy dependence of
the $e^+e^- \to f$ cross section. The first experimental study of the 
interference near the $\psi(3770)$ resonance was performed for 
the $e^+e^-\to p\bar{p}$ process in the BESIII experiment~\cite{ppbar}. 
In this work existing
data on the $e^+e^-\to K^+K^-$ and $e^+e^-\to K_SK_L$ cross sections
are used to study the interference near the $\psi(3770)$ and $\psi(4160)$ 
resonances and measure the cross sections for the resonant processes
$e^+e^-\to \psi(3770),\psi(4160)\to K^+K^-$.

\section{\bf\boldmath Fit to the $e^+e^-\to K^+K^-$ cross section}
\begin{figure}
\includegraphics[width=0.4\textwidth]{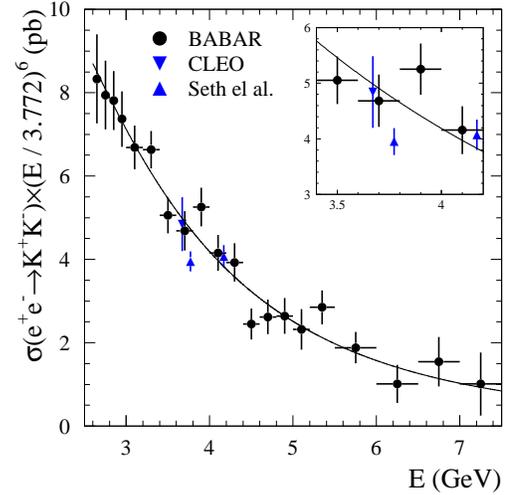}
\caption{The $e^+e^-\to K^+K^-$ cross section multiplied by the factor 
$(E({\rm GeV})/3.772)^6$ measured in the works: \cite{BABAR} (BABAR),
\cite{CLEO} (CLEO), and \cite{NU} (Seth {\it et al}). Data of 
Refs.~\cite{BABAR,CLEO} are approximated by Eq.~(\ref{eq1}).
The inset shows an enlarged version of the energy region near the
resonance $\psi(3770)$.
\label{fig1}}
\end{figure}
\begin{figure*}
\includegraphics[width=.4\textwidth]{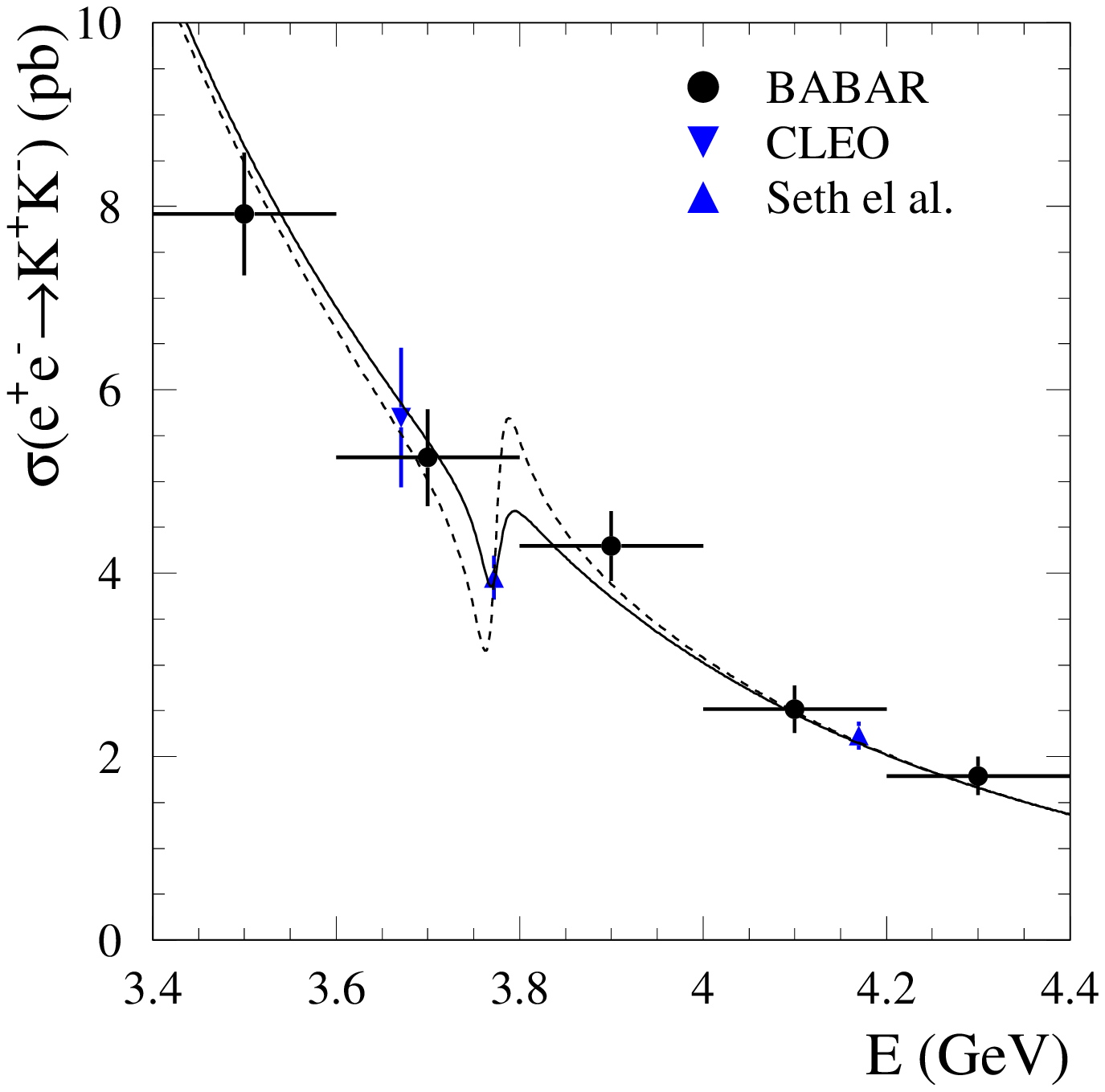}
\includegraphics[width=.4\textwidth]{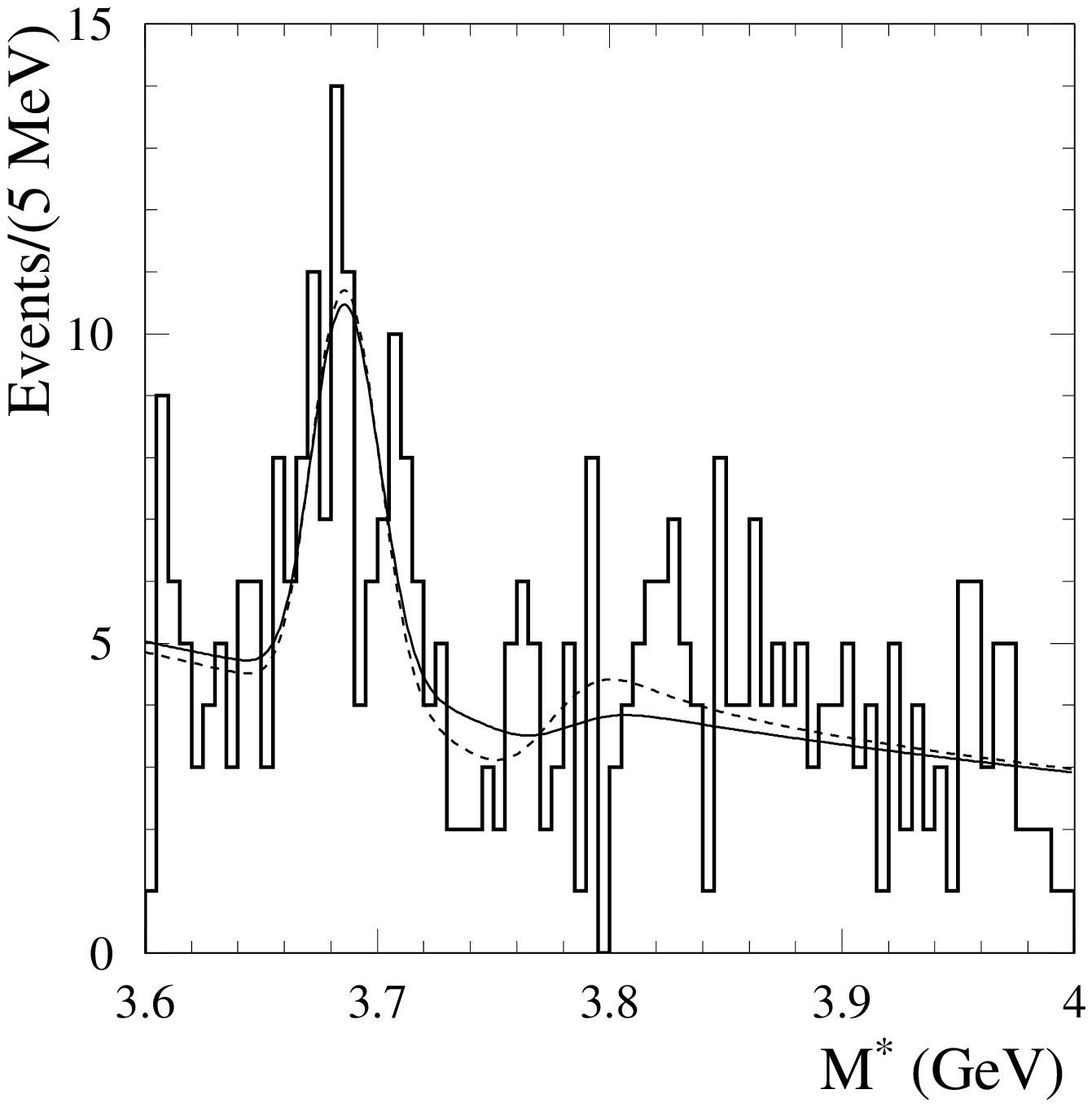}
\caption{Left: The $e^+e^-\to K^+K^-$ cross section near the $\psi(3770)$
resonance. Data are from Refs.~\cite{BABAR} (BABAR),
\cite{CLEO} (CLEO), \cite{NU} (Seth {\it et al}). Right:
The $K^+K^-$ invariant mass spectrum for the ISR process
$e^+e^-\to K^+K^-\gamma$ near the $\psi(3770)$ resonance
obtained in Ref.~\cite{BABAR}. 
 The dashed curves are the result of the fit
to the $e^+e^-\to K^+K^-$ data described in the text. The solid
curves are the result of the same fit with additional constraints from
the $e^+e^-\to K_SK_L$ data.\label{fig2}}
\end{figure*}
For the $e^+e^-\to K^+K^-$ process, we use the BABAR measurements in the 
center-of-mass energy region $E=2.6-7.5$ GeV obtained
using the initial-state radiation (ISR) method~\cite{BABAR}, and
direct measurements~\cite{CLEO,NU} at $E=3.671$, 3.772 É 4.17 çÜ÷
based on data collected in the CLEO experiment.
The measured energy dependence
of the $e^+e^-\to K^+K^-$ cross section above 2.6 GeV is shown 
in Fig.~\ref{fig1}. The curve is the result of the fit 
to the cross-section data with the function proposed 
in Ref.~\cite{BABAR}:
\begin{eqnarray}
\sigma_{\rm cont}(E)& = &\frac{\pi\alpha^{2}\beta^3}{3E^2}
|F_K(E)|^{2},\nonumber \\ 
|F_K(E)|&=&\frac{A}{E^2(E^\gamma+B)},
\label{eq1}
\end{eqnarray}
where $\alpha$ is the fine-structure constant,
$\beta =\sqrt{1-4m_K^2/E^2}$, $m_K^2$ is the charged kaon mass,
$A$, $B$, and $\gamma$ are fitted parameters. 
The measurements from Ref.~\cite{NU} performed near the maxima of the 
$\psi(3770)$ and $\psi(4160)$ resonances, are not included in the fit.

As seen in the inset, the point at 4.17 GeV is consistent with 
the approximation of the nonresonant cross section, whereas the 
point at 3.772 GeV lies about three standard deviations below.
The deviation may be a result of interference between the resonant and 
nonresonant amplitudes of the $e^+e^-\to K^+K^-$ reaction.

The cross section near the $\psi(3770)$ resonance is described by the
following formula:
\begin{eqnarray}
\sigma_{K^+K^-}(E)=
\left|\sqrt{\sigma_{\rm cont}}-\sqrt{\sigma_\psi}e^{i\phi}
\frac{m_\psi\Gamma_\psi}{D}\right|^2=\nonumber\\
\sigma_{\rm cont}+\left (\sigma_\psi+
2\sqrt{\sigma_{\rm cont}{\sigma_\psi}}\sin{\phi}\right)
\frac{m^2_\psi\Gamma^2_\psi}{|D|^2}-\nonumber\\
2\sqrt{\sigma_{\rm cont}\sigma_\psi}\cos{\phi}
\frac{m_\psi\Gamma_\psi(m^2_\psi-E^2)}{|D|^2},
\label{eq2}
\end{eqnarray}
where $\sigma_\psi\equiv \sigma_{\psi(3770)}$ is the cross section for the process 
$e^+e^-\to\psi(3770)\to K^+K^-$ in the resonance maximum,
$\phi$ is the relative phase between resonant and nonresonant
amplitudes, $D=m^2_\psi-E^2-im_\psi\Gamma_\psi$, and $m_\psi$ and $\Gamma_\psi$
are the $\psi(3770)$ mass and width, respectively.

Data from Refs.~\cite{BABAR,CLEO} and the measurement at $E=3.772$ GeV from
Ref.~\cite{NU} are fitted by the formula~(\ref{eq2}).
The measurements in Ref.~\cite{BABAR} were made using the ISR method. Therefore,
they are compared with the average
cross-section values over the corresponding energy intervals.
For the energy intervals near the $\psi(3770)$ resonance,
3.6--3.8 and 3.8--4.0 GeV, where the $e^+e^-\to K^+K^-$ cross section 
changes rapidly due to interference of resonant and nonresonant amplitudes,
the $K^+K^-$ invariant mass spectrum for the ISR process 
$e^+e^-\to K^+K^-\gamma$~\cite{BABAR} is used instead
of the cross section. The mass spectrum is described as follows 
\begin{eqnarray}
\frac{dN}{dM^\ast}&=&
\int R(M^\ast,M)\frac{dN}{dM}(M)dM-
\left ( \frac{dN}{dM^\ast} \right)_{\rm bkg},\\
\frac{dN}{dM}&=&
\sigma_{K^+K^-}(M)\frac{dL}{dM}(M)\varepsilon(M),
\end{eqnarray}
where $M$ and $M^\ast$ are the true and measured $K^+K^-$ invariant masses,
respectively, $R(M^\ast,M)$ is a function describing detector
mass resolution~\cite{BABAR}, 
${dL}/{dM}$ is the ISR luminosity (see, for example Ref.~\cite{BABAR}),
$\varepsilon$ is the detection efficiency, $(dN/dM^\ast)_{\rm bkg}$ is 
the mass spectrum of background events. The mass dependence of the ISR 
luminosity, detection efficiency and $(dN/dE_{\rm meas})_{\rm bkg}$ are 
obtained by interpolation between the values given in Ref.~\cite{BABAR} for 
mass intervals shown in Fig.~\ref{fig1}. The experimental mass spectrum 
contains also events from the decay $\psi(2S)\to K^+K^-$. The $\psi(2S)$ 
contribution is added to the fit with a shape
described by convolution of a Breit-Wigner resonance line-shape with the 
resolution function.
The measured cross and the mass spectrum are fitted simultaneously. 

The systematic uncertainty of the cross section measured in Ref.~\cite{BABAR} 
is separated into two parts. The first includes systematic errors of 
statistical origin, mainly due to background subtraction. This uncertainty
is added in quadrature to the statistical error of the cross section. 
The second part includes correlated uncertainties 
due to the data-MC simulation difference in the detection efficiency
and luminosity determination. This uncertainty ($\sigma_S$) is practically 
independent of energy and is equal to 2.4\%. In the fit, it is taken into 
account by multiplying the theoretical cross section [Eq.~(\ref{eq2})] for the 
BABAR measurements by a free parameter $S$, and by adding to the
logarithmic likelihood function the term $(S-1)^2/(2\sigma_S^2)$, which
allows all theoretical values for the BABAR measurements to be shifted
simultaneously inside $\sigma_S$. For the measurements of Refs.~\cite{CLEO,NU}
at $E=3.671$ and 3.772 GeV, the
statistical and systematic uncertainties are added in quadrature.

The fitted parameters are $\sigma_\psi$, $\phi$, the value of the nonresonant 
cross section at $E=3.772$ GeV, $\gamma$ and $B$ from Eq.~(\ref{eq1}),
the number of events from the $\psi(2S)\to K^+K^-$ decay, and $S$.
The result of the fit is shown in Fig.~\ref{fig2} by dashed curves.
The statistical significance of the $\psi(3770)\to K^+K^-$ decay is
calculated from the differences of the likelihood function values 
for the fits with free $\sigma_\psi$ and $\sigma_\psi=0$ and is 
found to be $3.2\sigma$.

\begin{figure}
\includegraphics[width=.4\textwidth]{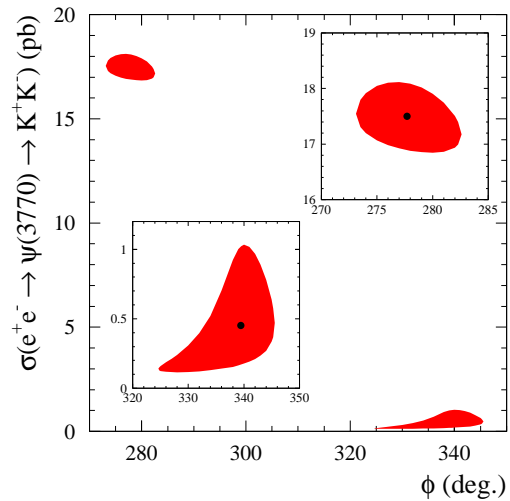}
\caption{The fitted value of the
$e^+e^-\to \psi(3770)\to K^+K^-$ cross section versus the
fitted value of the phase between resonant and nonresonant amplitudes
for the $e^+e^-\to K^+K^-$ reactions. The $1\sigma$ contours are
shown for the two solutions. The filled circles indicate 
the values corresponding to the minimum of the likelihood function.
\label{fig3}}
\end{figure}

The fit yields two solutions. They correspond to the same
values of the factors 
$\sigma_\psi+2\sqrt{\sigma_{\rm cont}{\sigma_\psi}}\sin{\phi}$
and $2\sqrt{\sigma_{\rm cont}\sigma_\psi}\cos{\varphi}$ in Eq.~(\ref{eq2}),
but different values of $\sigma_\psi$ and $\varphi$. The $1\sigma$
contours for these solutions are shown in Fig.~\ref{fig3}.

To determine the branching fraction the fitted cross section is
divided by the $e^+e^-\to \psi(3770)$ cross section,  which is
calculated as 
$\sigma_0=(12\pi/m_\psi^2)(\Gamma(\psi(3770)\to e^+e^-)/\Gamma_\psi)$. 
Unfortunately, the experimental situation with $\psi(3770)$ electronic width
is somewhat uncertain. The Particle Data Group~\cite{pdg} value 
$\Gamma(\psi(3770)=0.262\pm0.018$ keV corresponds to 
$\sigma_0\approx 9.9\pm0.8$ nb, which is significantly higher than
the value of the $e^+e^-\to D\bar{D}$ cross section 
$(6.57\pm0.04\pm0.10)$ nb measured by the CLEO Collaboration in the maximum 
of the $\psi(3770)$ resonance~\cite{ddbar}. Interference between resonant 
and nonresonant amplitudes in the $e^+e^-\to D\bar{D}$ reaction, which was
ignored in most $\Gamma(\psi(3770)\to e^+e^-)$ measurements, is a source
of additional uncertainty.
The analysis performed by the KEDR collaboration~\cite{KEDR} shows that taking
into account the interference decreases $\Gamma(\psi(3770)\to e^+e^-)$ by about
40\% compared with the value obtained ignoring the interference. In this paper
we will estimate the $\psi(3770)\to K^+K^-$ branching fraction using the value
$\sigma_0=6.36\pm0.08^{+0.41}_{-0.30}$ nb obtained by CLEO~\cite{gee}. The
close value was used previously in the measurements of 
$B(\psi(3770)\to\phi\eta)$~\cite{phieta}
and $B(\psi(3770)\to K_SK_L$~\cite{kskl}. 

The two obtained solutions correspond to the branching fractions
of about $10^{-4}$ and $3\times10^{-3}$. The latter significantly, more 
than by an order of magnitude, exceeds theoretical predictions~\cite{ddloop}.

\section{\bf\boldmath Constraints from $e^+e^-\to K_SK_L$ measurements}
Additional constraints on the $B(\psi(3770)\to K^+K^-)$ value can be
obtained from data on the $e^+e^-\to K_SK_L$ process. The branching 
fractions of $\psi(3770)\to K^+K^-$ and $\psi(3770)\to K_SK_L$ may
be different only due to single-photon contributions, which are related to
the values of the nonresonant $e^+e^-\to K^+K^-$ and $e^+e^-\to K_SK_L$ cross
sections. The cross section for the single-photon transition
$e^+e^-\to\psi(3770)\to \gamma^\ast \to K^+K^-$ is 
calculated as~\cite{gatto}
\begin{equation}
\sigma_{\psi,\gamma}=
\frac{9B(\psi(3770)\to e^+e^-)^2}{\alpha^2}\sigma_{\rm cont}(m_\psi)
\end{equation}
and is about $3\times10^{-5}$ pb. The corresponding branching fraction
is about $0.5\times 10^{-8}$. For the $K_SK_L$ final state, the single-photon
branching fraction is expected to be at least an order of 
magnitude smaller (see discussion below). Taking into account interference 
between electromagnetic and strong decay amplitudes, we conclude that for
$B(\psi(3770)\to K^+K^-)$ higher than $10^{-6}$ the single-photon 
contribution is negligible. Therefore, we expect that
$B(\psi(3770)\to K^+K^-)=B(\psi(3770)\to K_SK_L)$ is good approximation.

For the $e^+e^-\to K_SK_L$ process, there is an upper limit on 
the cross section at 3.773 GeV~\cite{kskl}, 
$\sigma_{K_SK_L}(m_\psi)<0.07$ pb at 90\% CL.
In Ref.~\cite{kskl} this value was used to obtain the upper limit
on $B(\psi(3770)\to K_SK_L)$. This approach,
however, does not take into account interference between
resonant and nonresonant amplitudes of the $e^+e^-\to K_SK_L$ process.
Data on the nonresonant cross section in the energy region of
interest are practically absent. There are two $e^+e^-\to K_SK_L$
measurements~\cite{dm1,KSKLbabar} near 2 GeV. Comparing these measurements
with data on the $e^+e^-\to K^+K^-$ cross section~\cite{KKbabar} we
estimate that $r=\sigma_{K_SK_L}/\sigma_{K^+K^-}=0.098\pm0.060$ at $E=2$ GeV.
At higher energy, there is only one measurement of this ratio at 4.17 GeV
$r=0.0144\pm0.0072$~\cite{KSKLNU}, which may be distorted by resonance 
contributions from the $\psi(4160)\to K\bar{K}$ decays. The theoretical 
prediction for this ratio obtained using leading-order leading-twist QCD calculation
of the kaon electromagnetic form factors, is $r\approx 0.04$~\cite{chernyak} 
in the energy region 3--4 GeV. 

To take into account the $e^+e^-\to K_SK_L$ data, we include in the fit,
described in the previous section, two additional measurements:
$\sigma_{K_SK_L}(m_\psi)=0.0\pm0.5$, which corresponds to the upper
limit $\sigma_{K_SK_L}(m_\psi)<0.07$ pb at 90\% CL, 
and $r(m_\psi)=0.030\pm0.15$, obtained from a linear approximation
between the $r$ values at $E=2$ and 4.17 GeV. The energy 
dependence of the $e^+e^-\to K_SK_L$ cross section near $\psi(3770)$ 
resonance is described by Eq.~(\ref{eq2}) with the replacement
of $\sigma_{\rm cont}(E)$ by $r(m_\psi)\sigma_{\rm cont}(E)$.
It is expected that in the energy region under study the nonresonant $K^+K^-$
and $K_SK_L$ amplitudes have the same sign of the real parts
and similar ratios of imaginary to real parts~\cite{chernyak_sign}.
Therefore, we assume that the phase $\phi$ is the same
for $e^+e^-\to K_SK_L$ and $e^+e^-\to K^+K^-$.

The fit with the $K_SK_L$ data yields a single solution:
\begin{eqnarray}
\sigma_{\psi(3770)}&=&0.073^{+0.62}_{-0.44}\mbox{ pb},\\
\phi&=&(308^{+17}_{-34})^\circ.
\end{eqnarray}
The $\sigma_{\psi}$ and $\phi$ values obtained with the $K_SK_L$ constraints are
shifted from unconstrained values obtained in the previous section by
about $1.5\sigma$. The statistical significant of the result is the
same, $3.2\sigma$. The fitted energy dependence of the 
$e^+e^-\to K^+K^-$ cross section and the fitted mass spectrum 
are shown in Fig.~\ref{fig2} by the solid curves.

\begin{figure}
\centering
\includegraphics[width=0.4\textwidth]{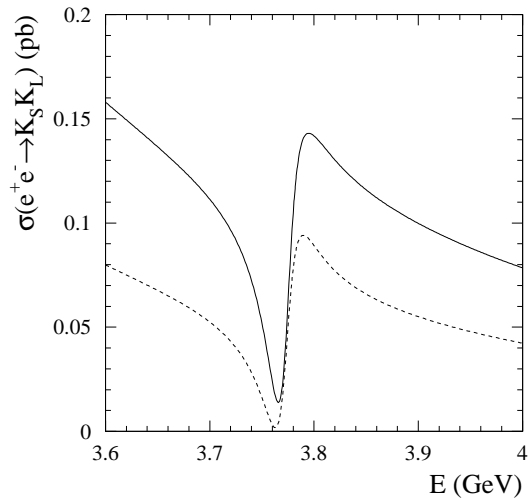}
\caption{The expected from the fit energy dependence of the 
$e^+e^-\to K_SK_L$ cross section near the $\psi(3770)$ resonance.
The solid and dashed curves correspond to different values of
the nonresonant $e^+e^-\to K_SK_L$ cross section at $E=m_\psi$,
0.117 pb and 0.061 pb, respectively.
\label{fig4}}
\end{figure}

The fitted value of the nonresonant $e^+e^-\to K_SK_L$ cross section
at $E=m_\psi$ is $0.117\pm 0.062$ pb. The expected energy dependence of
the $e^+e^-\to K_SK_L$ cross section near the $\psi(3770)$ resonance
is shown in Fig.~\ref{fig4} by the solid curve. To study dependence
of the result on the value of $r(m_\psi)$, the fit is performed
with the $r$ values obtained at 2 GeV and 4.17 GeV,
$r=0.098\pm0.098$ and $r=0.0144\pm0.0144$, respectively, taken with  
a 100\% uncertainty. The results for $\sigma_\psi$ and $\phi$ are changed 
insignificantly, while the value of the nonresonant $e^+e^-\to K_SK_L$ 
cross section
decreases to $0.061\pm0.061$ pb. The energy dependence of the cross
section obtained with modified $r(m_\psi)$ is shown in Fig.~\ref{fig4} 
by the dashed curve.

The branching fraction 
$B(\psi(3770)\to K^+K^-)$ corresponding to the measured value of the 
resonant cross section is about $10^{-5}$. It is an order of magnitude 
lower than the prediction~\cite{ddloop}, but two orders larger than 
the estimation not taking into account effects of intermediate $D\bar{D}$ 
loops [Eq.~(\ref{eq0})].

\section{\bf\boldmath Upper limit on the $\psi(4160)\to K^+K^-$ decay}
The fitting procedure described above is used in the energy region
of $\psi(4160)$ resonance. The BABAR~\cite{BABAR} and
CLEO~\cite{CLEO} nonresonant $e^+e^-\to K^+K^-$ data, and the measurement of 
the $e^+e^-\to K^+K^-$ cross section at $E=4.17$ GeV~\cite{NU} 
(see Fig.~\ref{fig1}) are fitted together with
the the $e^+e^-\to K_SK_L$ cross section measurement, 
$\sigma_{K_SK_L}=0.032\pm0.017$ pb at $E=4.17$ GeV~\cite{KSKLNU}.
To estimate the nonresonant $e^+e^-\to K_SK_L$ cross section, the value of 
$r=0.098\pm0.098$ is used in the fit.
The fitted value of the $e^+e^-\to \psi(4160)\to K^+K^-$ cross section in
the resonance maximum is found to be
$\sigma_{\psi(4160)}=0.006^{+0.047}_{-0.006}$. The corresponding upper limit
is
\begin{eqnarray}
\sigma_{\psi(4160)} < 0.062\mbox{ pb at 90\% CL}.
\end{eqnarray}
The $e^+e^-\to \psi(4160)$ cross section in the resonance maximum 
calculated from $B(\psi(4160)\to e^+e^-)=(6.9\pm3.3)\times10^{-6}$~\cite{pdg}
is equal to $5.8\pm2.8$ nb. Taking into account the uncertainty 
of the $\psi(4160)$ production cross section we estimate 
that $B(\psi(3770)\to K^+K^-)<2\times10^{-5}$ at 90\% CL.

\section{Summary}
Due to the relatively large continuum cross section for the $e^+e^-\to K^+K^-$
process, the $\psi(3770)\to K^+K^-$ and $\psi(4160)\to K^+K^-$ decays reveal 
themselves as interference patterns in the cross section energy dependence 
near the resonances. In this work, existing data on the $e^+e^-\to K^+K^-$ 
and $e^+e^-\to K_SK_L$ cross sections~\cite{BABAR,CLEO,NU,kskl,dm1,KSKLbabar,
KKbabar,KSKLNU} have been analyzed to obtain the interference parameters.  
For $\psi(3770)\to K^+K^-$ decay, the $e^+e^-\to \psi(3770) \to K^+K^-$ 
cross section in the resonance maximum and the interference phase 
are found to be 
\begin{eqnarray}
\sigma_{\psi(3770)}&=&0.073^{+0.62}_{-0.44}\mbox{ pb},\\
\phi&=&(308^{+17}_{-34})^\circ.
\end{eqnarray}
with a statistical significance of 3.2$\sigma$.
For the $\psi(4160)\to K^+K^-$ decay, the upper limit on
the cross section in the resonance maximum has been obtained:
\begin{eqnarray}
\sigma_{\psi(4160)} < 0.062\mbox{ pb at 90\% CL}.
\end{eqnarray}

\section{ \boldmath Acknowledgments}
The author thanks the BABAR Collaboration for providing data on 
the $K^+K^-$ invariant mass spectrum and the detector mass resolution 
near the $\psi(3770)$ resonance.
This work is supported by the Ministry of Education and Science of the 
Russian Federation.

\end{document}